# Performance Comparison and Analysis of Preemptive-DSR and TORA

V.Ramesh [1]  Dr.P.Subbaiah [2]  N.Koteswar Rao [3]  N.Subhashini [4]  Narayana.D [5]

[1] Research scholar, Sathyabama University, Chennai, TamilNadu, India.
`v2ramesh634@yahoo.co.in`

[2] Principal, Veerabrahmendra Institute of Technology & Sciences, Badvel, Kadapa-Dt, AP, India.
`subbaiah_nani@sify.com`.

[3] Asst.Professor, BVRIT, Narsapur, Medak, AP, India.
`subbu_narra@yahoo.com`.

[4] Assoc.Professor, Dept of IT, Narayana Engg. College, Gudur, AP, India.
`rao0007@gmail.com`.

[5] Asst.Professor, CSE, CMRCET, Hyderabad, AP, India.
`Narayanarao515@yahoo.com`

## *ABSTRACT*

*The Dynamic Source Routing protocol (DSR) is a simple and efficient routing protocol designed specifically for use in multi-hop wireless ad hoc networks of mobile nodes. Preemptive DSR(PDSR) is the modified version of DSR. The main objective of this paper is to analyze and compare the performance of Preemptive DSR and Temporarily Ordered Routing Algorithm(TORA).It discusses the effect of variation in number of nodes and average speed on protocol performance. Simulation results (provided by the instructor) are analyzed to get an insight into the operation of TORA and PDSR in small/large sized networks with slow/fast moving nodes. Results show that PDSR outperforms TORA in terms of the number of MANET control packets used to maintain/erase routes. Also, it is concluded that TORA is a better choice than PDSR for fast moving highly connected set of nodes. It is also observed that DSR provides better data throughput than TORA and that routes can be created faster in PDSR than in TORA. This paper tries to explain the reasons behind the nature of the results.*

## 1.INTRODUCTION

A mobile *ad hoc* network (MANET) is a self-configuring network of mobile devices connected by wireless links. In a MANET, each node acts as a router to establish end-to-end connections, and because the network topology between sources and destinations frequently changes, it is difficult to maintain and restore a route. To deal with these issues, several routing protocols for MANETs have been proposed.

In a MANET, routing protocols can be divided into two major categories - proactive routing and reactive routing. Proactive routing protocols, also known as table-driven routing protocols, contain information on every node and update the routing table information periodically. In DSDV, OLSR, and WRP, typical proactive routing protocols, the source node is equipped beforehand with information appertaining to the pathway of the destination node before it send data packets there. As a fatal drawback of the proactive routing approach, however, a mass of route-updating messages flood the entire network periodically to maintain the route information fresh. Furthermore, each node unnecessarily stores the full set of route information, especially in a highly mobile environment where the routing table of a node is updated frequently for dynamic topology. Each node must find the latest broadcast routing path information periodically. Such periodic updates cause unnecessary network overhead.

Reactive routing protocols, also known as on-demand routing protocols, do not conserve the routing table information; instead, they execute a route discovery procedure to determine a route to the destination only when the source node requires a path to the destination node. Once a route has been discovered, the route is maintained until the destination becomes inaccessible or the route is no longer desired. AODV, DSR and the Temporally Ordered Routing Algorithm (TORA) are representative examples of reactive protocols. Particularly with a large number of nodes, reactive routing protocols are more appropriate than a proactive routing approach.

## 2. Overview of PDSR and TORA

### 2.1 DSR[1]

The Dynamic Source Routing protocol is a simple and efficient routing protocol designed specifically for use in multi-hop wireless ad hoc networks of mobile nodes. Using DSR, the network is completely self-organizing and self-configuring, requiring no existing network infrastructure or administration. Network nodes cooperate to forward packets for each other to allow communication over multiple "hops" between nodes not directly within wireless transmission range of one other. As nodes in the network move about or join or leave the network, and as wireless transmission conditions such as sources of interference change, all routing is automatically determined and maintained by the DSR routing protocol. Since the number or sequence of intermediate hops needed to reach any destination may change at any time, the resulting network topology may be quite rich and rapidly changing.

The DSR protocol is composed of two main mechanisms that work together to allow the discovery and maintenance of source routes in the ad hoc network:

*Route Discovery* is the mechanism by which a node S wishing to send a packet to a destination node D obtains a source route to D. Route Discovery is used only when S attempts to send a packet to D and does not already know a route to D.

*Route Maintenance* is the mechanism by which node S is able to detect, while using a source route to D, if the network topology has changed such that it can no longer use its route to D because a link along the route no longer works. When Route maintenance indicates a source route is broken, S can attempt to use any other route it happens to know to D, or it can invoke Route Discovery again to find a new route for subsequent packets to D. Route Maintenance for this route is used only when S is actually sending packets to D.

## 2.2 Preemptive DSR

**The Algorithm:**

*a) Route Discovery:*

Step 1: When a source node S wants to send a data, it broadcast the RREQ packet to its neighbor nodes.

Step 2: When an intermediate node on the route to the destination receives the RREQ packet, it appends its address to the route record in RREQ and re-broadcast the RREQ.

Step 3: When the destination node D receives the first RREQ packet, it starts a timer and collects RREQ packets from its neighbors until quantum q time expires.

Step 4: The destination node D finds the two (primary +Backup) best routes from the collected paths (Step 3) within the quantum q time.

Step 5: The destination node D sends RREP packet to the source node S by reversing (RREQ) packets which includes the two routes (Primary +Backup) for further communication.

*b) Route Monitoring:*

Step 1: Each intermediate node on the route starts monitoring the signal strength.

Step 2: If signal strength falls below the specified threshold T, it will send a warning message "Path likely to be disconnected", to the source node S.

*c) The Source node S Communicates with destination node D:*

Step 1: The source node S starts Communicating with destination node D using primary path.

Step 2: On receiving the warning message from the intermediate node, it starts ommunicating destination node D with the backup route also.

Step 3: If source node S receives the acknowledgement form the destination node D go to step 4 else step 5.

Step 4: Preemption, switch over from Primary to Backup route.

Step 5: Initiates Route Discovery Process.

## 2.3 Temporarily Ordered Routing Algorithm [2]

In TORA, each node i knows its own height and the height of each of its directly connected neighbors j. It marks each link as upstream or downstream based on whether the height of its neighbor is greater or less than its own height. Nodes are assigned heights based on their location with respect to the destination. Nodes closer to the destination have smaller heights than the nodes away from it. When a node gets a data packet, it always forwards it in the downstream direction. Thus packets find their way to the destination flowing down from tall nodes located far away from the destination to short nodes located near it.

Consider the case where a link breaks due to the movement of some node. The node that loses this link raises its own height by selecting a new reference level and broadcasts an update to all its neighbors. If a neighbor has no downstream links, it raises its height (by selecting the propagated reference level) and reverses the direction of its upstream links. If on the other hand, the neighbor has a downstream link, no action needs to be taken at this node as an alternate route is discovered. In other words, a new reference level propagates from the point of failure, outwards to nodes that have lost all routes to the destination.

In Figure 1.a, the link between nodes F and E fails. When F sees this, it raises its height by choosing a new reference level. This reverses the link between D and F (1.b). The new reference level does not propagate any further as D has a downstream link to C. In Figure 1.c, the only link that connects the rest of the nodes to the destination fails. The new reference level propagates to all nodes in this case.

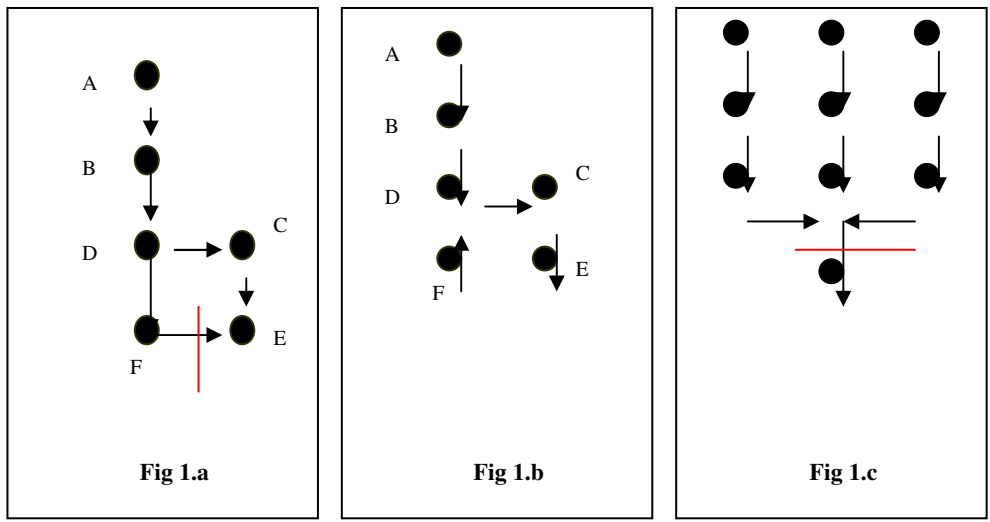

**Fig 1.a**   **Fig 1.b**   **Fig 1.c**

**Figure 1**

When a newly selected reference level propagates from the point of failure to the edge node, it selects a reference level higher than the propagated level. This is done by setting the reflect bit of the received reference level. This reflected level now propagates back towards the original source of failure. If the origin node of the reference level receives a reflected message from each one of its neighbors, no alternate routes exists and the routes must be deleted. A clear (CLR) message is propagated to clear existing routes in this case. But if the origin node does not receive a reflected level from even one of its neighbors, an alternate route exists and the CLR message is not sent. This is illustrated in Figure 2.

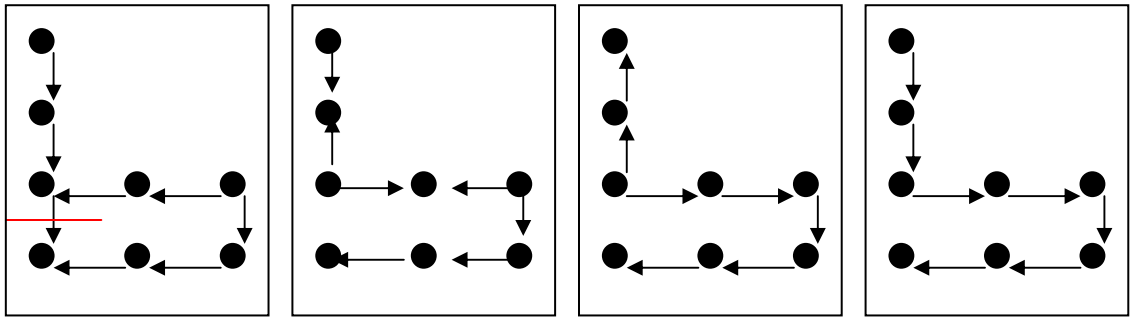

**Figure 2**

In the beginning of this discussion, it was assumed that nodes know their heights. But how are heights assigned to the nodes so that the destination has the smallest height and the node furthest away is the tallest. When a node wants to discover a route to the destination, it broadcasts a Query (QRY) packet to all its neighbors. If the neighbors don't know the height of the destination, they forward the QRY packet to their neighbors. If on the other hand, a node knows the height of the destination, it broadcasts an update (UPD) packet. On receiving an UPD packet, nodes learn the reference level of the destination and set their height to the same reference level but greater offset. This way as the UPD packet propagates outwards from the destination, the offset keeps increasing.

How fast the broadcast messages propagate through the network depends on the connectivity of nodes and the MAC layer used. For example consider the nodes in Figure 3. Consider Figure 3.a first. A wants to reach C and broadcasts a Query. B receives this query, waits for the medium to become free and transmits the Query to C. In Fig. 3.b, A sends Query to B,D and E. Since B,D and E share the same medium, B must wait for the other two to transmit before it gets a chance. The presence of highly connected nodes slows down the message propagation.

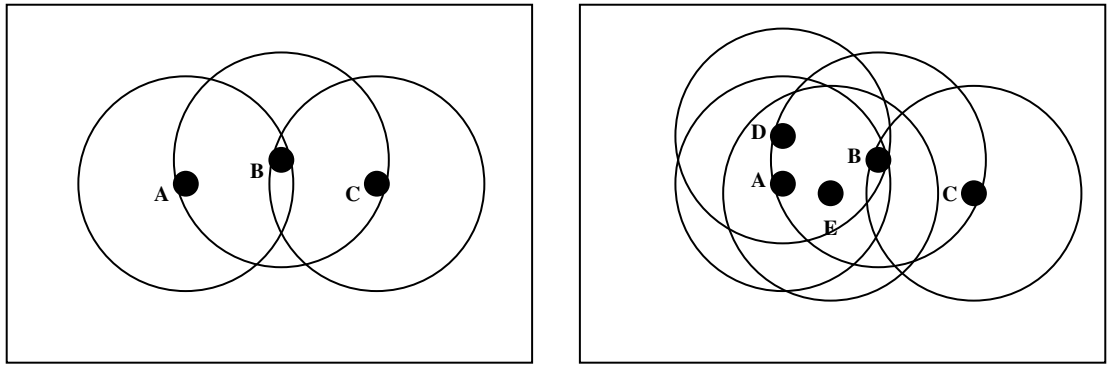

**Figure 3**

## 3. Discussion of Results

Table 1 below summarizes the simulation results provided by the instructor. Row labels indicate routing algorithm modeled, relative speed of users and number of stations. Column labels represent the performance metrics used in this study. In the simulation model, node 1 generates data to be sent to node 0. Other nodes don't generate any data and behave as simple MANET routers that generate/forward control packets and route data packets. The performance metrics Throughput (Data bits received by node 0 per unit time), % Packets Delivered (% of all data packets generated by node 1 that make it to the destination: node 0), Receiving Efficiency at node 0 (Bytes of data packets received / bytes of data packets received + bytes of MANET packets received), Sending efficiency at node 1 (Bytes of data packets successfully sent / (bytes of data packets successfully sent + bytes of MANET packets successfully sent and/or forwarded), Sending/Forwarding Efficiency in the Network (Same as sending efficiency but counts bytes of data packets successfully sent and/or forwarded) are used to study the effects of mobility and size and to compare the two protocols. The first part of this section discusses the results relative to the features discussed in section 1. Second part compares strength's and weaknesses of the two protocols. Note that the simulation results don't necessarily represent a trend, as they are an outcome of a single simulation run and not an average of results from multiple runs with different mobility patterns.

|  | Throughput (kBps) | % Packets Delivered | Receiving Efficiency (node 0) | Sending Efficiency (node 1) | Sending/Fwd Efficiency (network) |
|---|---|---|---|---|---|
| TORA Slow 10 | 178.40627 | 44.43422 | 100.00000 | 99.72400 | 98.99006 |
| TORA Fast 10 | 0.00000 | 0.00000 | NM | NM | NM |
| TORA Slow 30 | 4.11277 | 1.18765 | 100.00000 | 85.10018 | 25.22720 |
| TORA Fast 30 | 245.70120 | 60.23327 | 100.00000 | 99.66244 | 95.86307 |
| PDSR Slow 10 | 383.37242 | 94.10000 | 99.99890 | 99.99640 | 99.98251 |
| PDSR Fast 10 | 417.56675 | 99.66668 | 99.99987 | 99.99896 | 99.99347 |
| PDSR Slow 30 | 425.21003 | 99.99598 | 99.99991 | 99.99991 | 99.99977 |
| PDSR Fast 30 | 392.14352 | 93.86772 | 99.99954 | 99.99716 | 99.97648 |

**Table 1**

## 3.1 TORA

In the 10-node case nodes are sparsely connected and there is a high likelihood of the network being partitioned. For nodes using TORA, data throughput is 177.40627 kBps and 43 percent of data packets reach the destination. 90 percent of the lost packets are dropped at node 1. This happens when node 1 does not have any downstream links. About 10 percent of lost packets are dropped at intermediate nodes. This happens due to data transmissions during the time when a link fails and changes that reflect this failure are made in router states. Since the nodes are sparsely connected, alternate routes rarely exist. As routes must be cleared in this case, data packets transmitted during 'propagation/reflection of reference levels' and 'Clearing of routes' are lost.

When 30 nodes are placed in the same area, the probability of a partition decreases and the nodes are densely connected. For 30 fast moving nodes using TORA, data throughput is 245 kBps and 60 percent of the data packets reach the destination. Most packets are dropped at node 1. About 3-4 percent of the lost packets are lost in transit. This number is less than the 10-node case where 10 percent of all lost packets were lost in transit. This is because the nodes are densely connected when there are 30 nodes and there is a greater likelihood of finding an alternate path in a nearby node. If a node near the point of failure has an alternate route to the destination, TORA reacts very fast and not many packets are lost in transit. New reference levels do not propagate very far in this case. It does not take a lot of time to propagate these levels to nearby nodes and only a small number of packets that are transmitted during this time are lost.

It can be seen from the slow 30-node case that only a small percentage of packets generated by node 1 are delivered to node 0. This is probably because the network is partitioned and node 0 is unreachable from node 1 for a large portion of the total time. This does not represent a general trend, as for a large number of highly connected nodes with low mobility, network partitions should be rare. This seems to be a special case where node 0 cannot be reached.

As discussed in Section 1, many control packets are generated in the case of TORA. Since there are a large number of densely connected nodes, a control message broadcast by a node propagates through the entire network consuming Bandwidth (and power). Consider the case where the network becomes partitioned because the destination (node 0) moves away from all other nodes (Figure 4). New reference levels propagate outwards from the origin of error towards the node at the other end. Here the reference level is reflected towards the origin node. Again the reflected level propagates via all other nodes to the origin node. The origin node now generates a clear message that clears all routes. Most of the nodes involved in the control message transmission process never even transmit any data to node 0. First they expend bandwidth and power creating routes and then they erase these routes. This problem is illustrated by the parameter Sending/Forwarding Efficiency in the network (Table 1). The lower the value of Sending/Forwarding Efficiency in network, more is the bandwidth wasted in transmitting control messages.

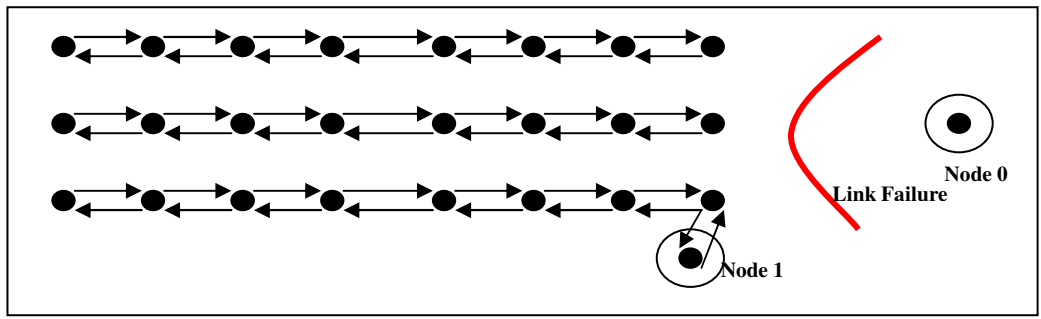

Figure 4

In the case of 10 fast moving nodes, node 1 never transmits a single data packet. The network is partitioned and it is not possible for node 1 to send data to node 0. Either the partition exists throughout the simulation or it joins and breaks before TORA can discover a route for node 1.

## 3.2 PDSR

It can be seen from the simulation results that data throughput is higher in PDSR than in TORA. This is due to the Data Salvage property of PDSR. When a link becomes bad, the PDSR node tries to find alternate paths in its local cache. If found, this path is used to salvage the data packet. This might also be because PDSR distributes its data traffic across all possible routes increasing data throughput.

Sending/Forwarding efficiency of the network is very high for PDSR. This is because Route request is the only control message that propagates through nodes that are not directly involved in routing data packets from node 1. Error message might be generated once in a while. Error messages propagate through the routing path and clear routes from the cache.

The data throughput is greater in the case of 30 slow moving nodes than in the case of 30 fast moving nodes. For fast moving nodes, it is more likely that the route stored in the local cache of some router is stale. In that case, the errors message propagates to the source and the source broadcasts a new route request message.

### 3.3 Strengths and Weaknesses of PDSR and TORA
- In densely connected networks, TORA control packets consume a considerable amount of bandwidth. Even if only one node is transmitting data, other nodes exchange some control packets. As one instance of TORA is running for each possible destination, number of control messages increase with an increase in the number of destinations.
- TORA recovers very fast from link failures in densely connected networks. In such networks, there are many alternate routes to the destination. A node that loses its downstream link selects a high reference level and propagates the update message. If any of the neighboring nodes have an alternate route, nothing needs to be done as reversing the link makes changes in routing path.
- PDSR performs very well as compared to TORA in terms of the number of control messages that must be exchanged to modify existing routes. So very little bandwidth and transmission power is consumed in this regard. But there is a source route header overhead that accompanies every packet. For smaller routes, this overhead is pretty low.
- PDSR creates new routes faster than TORA. In PDSR, the Routing response message is directed towards the source (assuming bi-directional links). In TORA on the other hand, UPD packet is broadcast. Broadcast packets take longer to reach a destination than directed messages. This was explained in section 1.
- In PDSR, it is difficult to maintain a route to the destination without expending a lot of bandwidth if the nodes are moving very fast. The routes stored in the routing cache become stale very fast and new route requests need to be generated for link failures.
- PDSR offers greater throughput than TORA. This is because more than one routes may be discovered at the same time when a route request is sent. Packets can be sent to destination using more than one route at the same time. This is not possible in TORA as there is only one route to the destination at any given time. [4]

## 4. Simulation Results

The simulations were performed using Network Simulator 2 (Ns-2), particularly popular in the ad hoc networking community. The traffic sources are CBR (continuous bit –rate). The source-destination pairs are spread randomly over the network. The mobility model uses 'random waypoint model' in a rectangular filed of 500m x 500m with 50 nodes. During the simulation, each node starts its journey from a random spot to a random chosen destination. Once the destination is reached, the node takes a rest period of time in second and another random destination is chosen after that pause time. This process repeats throughout the simulation, causing continuous changes in the topology of the underlying network. Different network scenario for different number of nodes and pause times are generated. The model parameters that have been used in the following experiments are summarized in Table 2.

| Parameter | Value |
|---|---|
| Simulator | NS-2 |
| Protocols studied | AODV, DSR and TORA |
| Simulation time | 200 sec |
| Simulation area | 500 x 500 |
| Transmission range | 250 m |
| Node movement model | Random waypoint |
| Traffic type | CBR (UDP) |
| Data payload | Bytes/packet |
| Bandwidth | 2 Mbps |

**Table 2: Simulation Parameters**

*Performance Indices*
The following performance metrics are considered for evaluation:
*Packet Delivery Fraction (PDF)*: The ratio of the data packets delivered to the destinations to those generated by the sources. Mathematically, it can be expressed as:

$$P = \frac{1}{c} \sum_{f=1}^{e} \frac{R_f}{N_f}$$

where P is the fraction of successfully delivered packets, C is the total number of flow or connections, f is the unique flow id serving as index, Rf is the count of packets received from flow f and Nf is the count of packets transmitted to f.

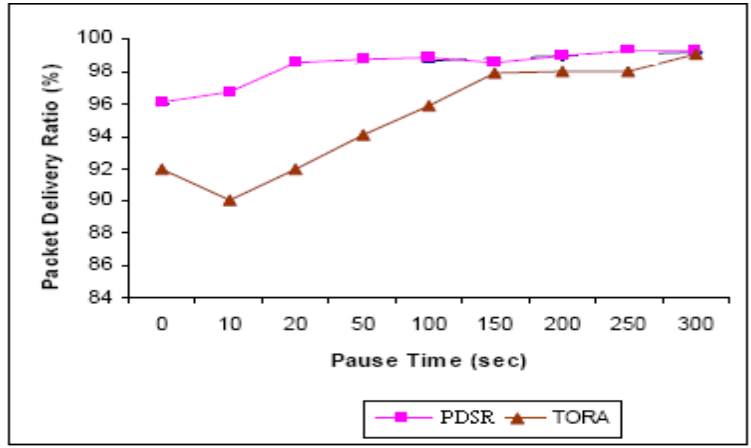

Figure 5: Packet Delivery Ratio Vs Pause Time for 50-node model with 10 sources

*Average end-to-end delay:* This includes all possible delays caused by buffering during route discovery latency, queuing at the interface queue, retransmission delays at the MAC, and propagation and transfer times. It can be defined as:

$$D = \frac{1}{N} \sum_{i=1}^{s} (r_i - s_i)$$

where N is the number of successfully received packets, i is unique packet identifier, ri is time at which a packet with unique id i is received, si is time at which a packet with unique id i is sent and D is measured in ms. It should be less for high performance.

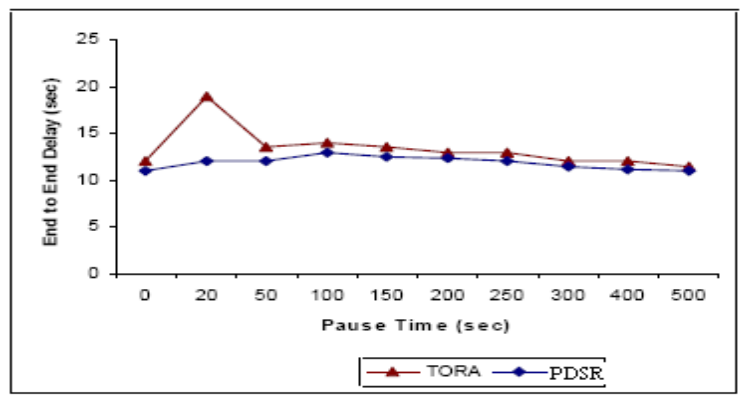

Figure 6: End to End delay Vs Pause Time 50-node model with 10-sources

## 5. Conclusions

Efficient data traffic routing is a very difficult problem in multi-hop wireless networks. Wireless nodes keep changing their position in an unpredictable manner. A routing algorithm must keep track of the destination's position and modify routes whenever the old routes become stale. In doing so, it must not generate a lot of control packets. This is important because wireless nodes are typically battery powered and are connected to each other via low bandwidth links. Power and bandwidth are precious resources and must be conserved wherever possible. Transmission of control packets uses up a bandwidth and should be minimized by designing routing algorithms in a smart way.

In this paper, we have studied the operation and performance of two popular routing algorithms: TORA and PDSR. It is concluded that PDSR outperforms TORA in terms of the control overhead. Nodes using TORA generate many control packets that use up precious power and bandwidth. PDSR is more efficient in this sense as it limits the number of control packets in the network. TORA outperforms PDSR for densely connected fast moving nodes. TORA reacts quickly to link failures and modifies existing routes. In PDSR, cached routes become stale very fast if the nodes change their positions rapidly. To create new routes, the source must generate a route request message.  Also, it is concluded that PDSR provides better data throughput than TORA. This is due to the ability of PDSR to acquire multiple routes in a single route request.

## ABOUT AUTHORS

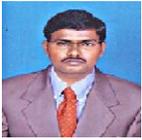

**1. Mr.V. Ramesh** received his B.Tech from N.B.K.R.I.S.T, Vidyanagar, AP in Computer Science & Engineering and M.Tech in IT from Sathyabama University, Chennai. Presently he is working as Associate Professor at Atmakur Engineering College, Atmakur, Hyderabad, AP. He has published several papers in various International & National Conferences and Journals. Presently he is pursuing his Ph. D in the field of Ad-hoc networks at Sathyabama University. His research interests include Operating Systems, Computer Networks and Data Mining.

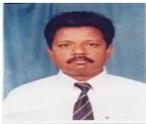

**2. Dr P. Subbaiah** received M.Tech(D.S.C) from JNTU and Ph. D from S.K University, Ananthapur in the area of fault tolerant systems. He has published several papers in international, national conferences and journals. He guided 6 research scholars. Presently he is working as Principal, Veerabrahmendra Institute of Tech & Science, Badvel, Kadapa, AP, India. His research interests include Mobile ad-hoc networks, Digital image processing and VLSI design.

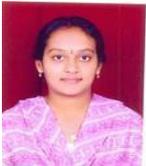

3**. Mrs N. Subhashini**, working as Asst.Professor in the department of CSE, B.V.Raju Institute of Technology & Science, Narsapur, Medak-Dt,AP, India.

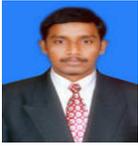 **4**. **Mr N. Koteswar Rao** received his B.Tach from Narayana Engineering College, Nellore and M.Tech from Sathyabama University, Chennai. Currently he is working as Assoc. Professor in the department of IT at Narayana Engineering College, Gudur, AP, India.

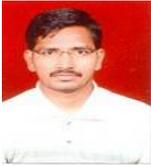 **5**. **Mr.D.Narayana,** received M.Tech from ANU, Guntur. He is working as Asst. Professor in the department of CSE at CMR College of Engg. & Tech., Hyderabad.